\documentclass[12pt,english]{iopart}
\usepackage[T1]{fontenc}
\usepackage[latin1]{inputenc}
\usepackage{graphicx}
\usepackage{setspace}
\usepackage{color}
\usepackage{amstext}
\usepackage{amssymb} 
\usepackage{cmap}
\usepackage{subfigure}

\makeatletter

\newcommand {\bv}{\mathbf{v}}

\newcommand {\bJ}{\mathbf{J}}

\newcommand {\bB}           {\mathbf{B}}

\newcommand {\PT}            {\overline{\mathbf{P}}}
\newcommand {\IT}            {\overline{\mathbf{I}}}

\newcommand {\para}		{\parallel}

\def\code#1{\texttt{#1}}

\usepackage{babel}
\makeatother
\begin{document}

\title[The impact of anisotropy on ITER scenarios]
{The impact of anisotropy on ITER scenarios} 

\author{M. J. Hole$^{1,2}$, Z. Qu$^{1}$, S. Pinches$^{3}$, M. Schneider$^{3}$, I. L. Arbina$^4$, M. J. Mantsinen$^{4,5}$, O. Sauter$^6$} 

\address{$^1$ Mathematical Sciences Institute, Australian National University, Acton 0200, ACT Australia}
\address{$^2$ Australian Nuclear Science and Technology Organisation, Locked Bag 2001, Kirrawee DC, NSW, 2232, Australia}
\address{$^3$ ITER Organization, Science Division, Route de Vinon-sur-Verdon, CS 90 046, 13067 St Paul-lez-Durance Cedex, France}
\address{$^4$ Barcelona Supercomputer Center (BSC), Barcelona, Spain}
\address{$^5$ ICREA, Barcelona, Spain}
\address{$^6$ Swiss Plasma Center, Switzerland}

\begin{abstract}
We report on the impact of anisotropy to tokamak plasma configuration and stability. Our focus is on analysis of the impact of anisotropy on ITER pre-fusion power operation 5~MA, $B=1.8$~T ICRH scenarios.  To model ITER scenarios remapping tools are developed to distinguish the impact of pressure anisotropy from the change in magnetic geometry caused by an anisotropy-modified current profile.  The remappings iterate the anisotropy-modified current profile to produce the same $q$ profile with matched thermal energy.  The analysis is a step toward equilibria that are kinetically self-consistent for a prescribed scenario. We find characteristic detachment of flux surfaces from pressure surfaces, and an outboard (inboard) shift of peak density for $T_{\para}>T_\perp$ ( $T_{\para}<T_\perp$). Differences in the poloidal current profile are evident, albeit not as pronounced as for the spherical tokamak.  We find that the incompressional continuum is largely unchanged in the presence of anisotropy, and the mode structure of gap modes is largely unchanged.  The compressional branch however exhibits significant differences in the continuum.  We report on the implication of these modifications.
\end{abstract}

 

\maketitle

\section{Introduction}

It is well known that the introduction of auxiliary  heating in tokamak plasmas can introduce toroidal and poloidal flows, and pressure anisotropy.  Studies in MAST suggest that the beam injected anisotropy is as much as $p_\perp/p_\para = 1.7$  \cite{Hole_anisotropy_11}, while in JET ICRH can cause anisotropies of $p_\perp/p_\para=2.5 $ \cite{Zwingmann_01}. Despite this the vast majority of plasma scenario codes solve the static isotropic Grad-Shafranov equation for two flux profiles and either a free or fixed boundary. 
The formulation of the anisotropic flowing equilibrium has been studied by many authors. \cite{Cooper_80, Salberta_1987, Iacono_90,Ilgisonis_1996, Guazzotto_04,Hole_Dennis_09, Fitzgerald_2013, Qu_2014, Gorelenkov_2018}
In one such formulation based on an enthalpy formulation \cite{Iacono_90}, the flow-modified MHD equations read
\begin{eqnarray}
\nabla \cdot (\rho \bv) & = & 0, \\
\rho \bv \cdot \nabla \bv & = & \bJ \times \bB - \nabla \cdot P  \\
\nabla \cdot \bB & = & 0, \\
\mu_0 \bJ  & = & \nabla \times \bB, \\
\nabla \times (\bv \times \bB) & = & 0
\end{eqnarray}
together with 
\begin{eqnarray}
\PT     & = & p_\perp \IT + \Delta \bB \bB / \mu_0, \\
\Delta & = & \frac{\mu_0 (p_\perp - p_\para)}{B^2}
\end{eqnarray}
If one assumes a frozen-in flux condition, neglects poloidal flow, assumes toroidal symmetry and assumes a two-temperature
Bi-Maxwellian model 
\begin{eqnarray}
p_\para & = & \frac{k_B \rho T_\para}{m}  \\
p_\perp & = & \frac{k_B \rho T_\perp}{m}  = \frac{k_B \rho T_\para}{m} \frac{B}{B - \Theta(\psi) T_\para} 
\end{eqnarray}
then the generalised Grad Shafranov equation becomes: 
{\setlength{\mathindent}{0cm}
\begin{equation}
\noindent \nabla \cdot \frac{(1-\Delta) \nabla \Psi}{R^2} = -\frac{F(\psi) F'(\psi)}{(1-\Delta) R^2} - 
\mu_0 \rho \left [ { T_\para '(\psi) + H'(\psi) + R^2 \Omega(\psi) \Omega'(\psi) - \left ( { \frac{\partial W}{\partial \psi} }  \right )_{\rho, B} } \right ] \label{eq:GS}
\end{equation}}
with five constraints: $\{ F(\psi), \Omega(\psi), H(\psi), T_\para (\psi), \Theta(\psi) \}$.   
Here, $\rho$ is the mass density, $\overline{m}$ the fluid mass, $k_B$ Boltzman's constant and $\mu_0$ the permeability of free space. 
In the enthalpy formulation of the generalised Grad Shafranov equation the flux function $H(\psi)$ is related to the enthalpy $W(\rho, B, \psi)$ and the toroidal rotation angular frequency $\Omega(\psi)$  through 
\begin{equation}
H(\psi) = W(\rho, B, \psi) - \frac{1}{2} \Omega^2 R^2. \label{eq:H}
\end{equation}
with 
\begin{equation}
W(\rho, B, \psi)  = T_\para \ln \frac{T_\para \rho}{T_\perp \rho_0},  \quad \quad \rho =  \rho_0 \frac{T_\perp}{T_\para} \exp \frac{H + \frac{1}{2} R^2 \Omega^2}{T_\para}  \label{eq:rho}
\end{equation}
This problem has been solved by 
the magnetic reconstruction code EFIT-TENSOR \cite{Fitzgerald_2013}, the fixed boundary code HELENA+ATF \cite{Qu_2014}, as well as the poloidal-flow enabled equilibrium code FLOW \cite{Guazzotto_04} and multi-specie code FLOW-M \cite{Hole_Dennis_09}. 
We have used these codes have been used to explore the impact of anisotropy and rotation in MAST. 

In Hole \etal \cite{Hole_anisotropy_11} we computed MAST equilibria with anisotropy and flow using FLOW, assuming 
a 100\% fast particle fraction. We showed that the unconstrained addition of anisotropy, obtained by modifying the pressure and rotation profiles, produced a change in $q_0$ from 1 to $ q_0 \approx 2.2$ . This is much larger than the change in $q$ profile due to the change in toroidal Alfv\'enic Mach flow from $M_{A,\phi}=0$ to $M_{A,\phi}=0.3$ . 
In \cite{Hole_2013} a proxy for the change in MHD stability for the same discharge was determined by using isotropic MHD stability codes CSCAS \cite{CSCAS_Poedts_93} and MISHKA \cite{MISHKA1_97} with an equilibrium for which the $q$ profile was remapped to match the $q$ profile of the anisotropic plasma. They found that reverse shear associated with anisotropy created a core localised odd TAE with harmonics of opposite poloidal mode number $m$.  Using the variational treatment of Smith, the CAE mode frequency was also computed for both isotropic and anisotropic plasmas, and the computed frequencies found to span the observed frequency range. 	The most extensive study of anisotropy is the treatment by Layden \etal \cite{Layden_16}, in which EFIT-TENSOR and EFIT++ reconstruction codes are used to find solutions with and without anisotropy, respectively. The safety factor profile in the isotropic reconstruction is reversed shear while
the anisotropic reconstruction gives monotonic shear; the isotropic TAE gap is much narrower than the anisotropic TAE gap; and the TAE radial mode structure is wider in the anisotropic case. These lead to a modification in the resonant regions of fast-ion phase space, and produce a 35\% larger linear growth rate and an 18\% smaller saturation amplitude for the TAE in the anisotropic analysis compared to the isotropic analysis.

The above studies showed that the variation of additional free parameters (anisotropy and flow) in general produce different magnetic configurations. 
Such variation is only meaningful if the differences are the result of different reconstruction codes that determine optimal fits to the data. 
In contrast, the unconstrained variation of input profiles $\{F(\psi), \Omega(\psi),H(\psi), T_\para (\psi), \Theta(\psi) \}$ can generate solutions that are either not fully self-consistent with an imposed constraint (like density), and/or  it is the inferred quantities (e.g. $ q$  profile) that are effectively constrained (by control of current profile).  For these cases a remapping choice must be made.  In \textit{this} work we implement at mapping that remaps the toroidal flux profile and pressure scaling to preserve the $q$ profile and stored energy, respectively.   The procedure enables a systematic study of the impact of toroidal flow and anisotropy for the same magnetic configuration ($q$ profile) and stored energy. 


\section{Equilibrium Mapping}

We have implemented remapping procedures by external iteration of \code{HELENA+ATF}, a fixed boundary solver for Eq. (\ref{eq:GS}).
 HELENA+ATF takes as input a boundary profile $(R,Z)$, five normalised flux profiles $\code{VF2} = \frac{F(\psi)^2}{F(0)^2}, \code{VTE}= \frac{T_\para (\psi)}{T_\para(0)}, \code{VOM2} = \frac{\Omega(\psi)^2}{\Omega(0)^2}, 
\code{VH} = \frac{H(\psi)}{H(0)}, \code{VTH} = \frac{\Theta(\psi)}{\Theta(0)}$, together with the scalars 
$\code{B} = \mu_0 \rho \frac{k_B a^2}{m_i \epsilon^2} \frac{T_\para(0)}{F(0)^2} = \beta(0)/2, \code{HOT} = \frac{m_i}{k_B} \frac{H(0)}{T_\para(0)}, 
\epsilon = a/R_0, R_0, B_0, \code{OMGOT} = \frac{m_i}{k_B} \frac{R_0^2 \Omega(\psi)^2}{T_\para(0)}$ and $\code{THTOF} = R_0 \Theta(\psi) \frac{T_\para}{F(0)}$. Here $F(0), T_\para (\psi), \Omega(0), H(0),$ and $\Theta(0)$ denote on-axis values, $a$ is the minor radius, $R_0$ and $B_0$ the geometric
axis and field strength at the geometric axis. Next, we outline the procedure for constraining a HELENA+ATF to a Grad-Shafranov solver, and adding the physics of anisotropy and flow. \\

\noindent 1. A commonly used file format for Grad-Shafranov solvers like EFIT is the \code{eqdsk} or \code{gfile} format, which supplies the plasma boundary, the toroidal flux $f(\psi)$ and thermal pressure $p(\psi)$ profiles.  The first step involves prescribing $F(\psi)^2/F(0)^2 = f(\psi)^2/f(0)^2$ and $T_\para(\psi)/T_\para(0) = p(\psi)/p(0)$. The three remaining flux functions \code{OM2, H} and \code{TH} can be arbitrarily set providing the scalar flags $\code{HOT}=\code{OMGOT} = \code{THTOF}=0$. The solution is fully constrained with the inclusion of $B_0$, the vacuum field strength at the plasma magnetic axis $R_0$, and the aspect ratio $\epsilon = a/R_0$, with $a$ the minor radius, and $\code{B}=p(0) \mu_0 \frac{R_0^2}{f(0)^2} = \beta(0)/2$. \\

\noindent 2. The addition of flow and anisotropy resolves the density profile, and so a choice must be made for $\rho(\psi) = m_i n_i(\psi)$  in the isotropic static limit.  If $n_i(\psi)$ and $T_\para(\psi)$ are consistent with an isotropic Grad-Shafranov solution, with  the thermal closure condition $p(\psi) \propto n_i(\psi) T_\para(\psi)$ then no further scaling for $T_\para(\psi)$ is required.
If however $p(\psi) \propto T_\para(\psi)$ as in step (1), then $T_\para(\psi)$ is rescaled
such that $T_\para(\psi) \propto p(\psi)/n_i(\psi)$ to preserve the pressure profile. 
To complete the specification of $H$ a value of $\rho_0$ is required.  We have selected $\rho_0 = \rho(0)/2$, which implies $H(0) \neq 0$, and places the sign change in $H(\psi)$ away from $\psi=0$.  The former is important as $H(0)$ is scaled in the code to be 1. Finally,  the parameter $\code{HOT} =  \frac{m_i}{k_B} \frac{H(0)}{T_\para(0)}$ is calculated. \\

\noindent 3. Nonzero anisotropy is specified either by an overall constant scaling or a profile. A constant scaling can be implemented through 
$\code{THTOF} = R_m \Theta(0) T_\para(0)/F(0) = (1-T_\para(0)/T_\perp(0))$ with  $\Theta(\psi)/\Theta(0)=1$.  A profile is constructed by taking
$\Theta(\psi)/\Theta(0)$ from the moments of an ICRH or NBI computed distribution function:
for off-axis peaks we have normalised the magnitude of $\Theta$ to its peak value  
$\Theta = \Theta/\Theta_{max}$, and renormalised  $ \code{THTOF}  \rightarrow \code{THTOF} \max (\code{VTH}) / VTH(0)$.
The parameter $\Theta(0)$ can be solved for $T_\perp(0)/T_\para(0)$ giving
$T_\perp(0)/T_\para(0) = 1/|1 - \Theta(0)|$ so that to lowest-order the stored energy will be 
preserved using $ \code{B} \rightarrow 3 \code{B} /(2 \times T_\perp(0)/T_\para(0)+ 1)$. Initialisation is completed using
$\code{HOT} \rightarrow \code{HOT} + log (1-\Theta(0))$.  \\

\noindent 4. We have undertaken two parameter scans to preserve the thermal energy to the isotropic value. First, HELENA+ATF equilibria are iterated with a simple shooting method for \code{B} until the stored thermal energy, $W_{thermal}$, matches the isotropic value.  Second, we have modified $F^2$ to achieve a match with either current or $q$ profile. The toroidal current can be written: 
\begin{equation}
J_\phi = -\frac{F(\psi) F'(\psi)}{(1 - \Delta ) R \mu_0}  - \rho \left [ {T'_\para(\psi) + H'(\psi) - \left ( \frac{\partial W}{\partial \psi} \right )_{\rho, B} } \right ]
\end{equation}
Next, we have assumed that 
\begin{equation}
J_{\phi,a} +\frac{F_a(\psi) F_a'(\psi)}{(1 - \Delta_a ) R \mu_0}  \approx 
J_{\phi,i} +\frac{F_i(\psi) F_i'(\psi)}{(1 - \Delta_i ) R \mu_0} 
\end{equation}
and computed
\begin{equation}
\int_1^{\psi_n} \langle J_{\phi,a} (1 - \alpha \Delta_a) R - J_{\phi,i} R \rangle \mu_0 (\psi_a - \psi_0)
\approx [ -F_a^2 + F_i^2] = - \delta F^2 (\psi) + \delta F^2.   \label{eq:J_update}
\end{equation}
We have implemented two choices of $\alpha$: $\alpha = 0$ is a prescription for the change in 
toroidal current, while $\alpha=1$ prescribes the change in $F$.  
Finally, the toroidal flux function is then updated 
\begin{equation}
F^2(\psi_n) \rightarrow F^2(\psi_n) - \lambda \delta F^2. 	\label{eq:F_update}
\end{equation}
where $\lambda$ is a relaxation parameter.  
We have computed the two metrics $\delta q$ and $\delta J_\phi$ given by 
\begin{eqnarray}
\Delta q & = & \int_0^1 (q_{target} - q)^2 d \psi_n  \\
\Delta J_\phi. & = & \int_0^1 (J_{\phi, i} - J_{\phi, a})^2 d \psi_n  
\end{eqnarray}
The remapping continues until either a target metric is met, $\delta F$ vanishes or a set iteration count is reached. 

\section{ITER scenario}

\begin{figure}[h]
\centering
\includegraphics[width=80mm]{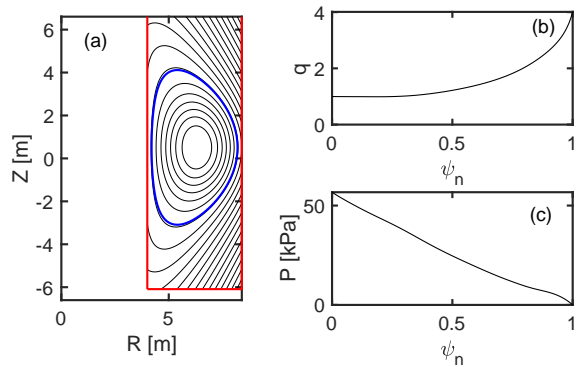}
\caption{\label{fig:ITER_scen1} 
ITER equilibrium scenario \#1000003 in the IMAS database.  Panel (a) shows the poloidal cross-section, and (b) and (c) the 
$q$ profile and pressure profile respectively. }
\end{figure}

We have demonstrated the equilibrium remapping technique for a pre-fusion power operation H plasma at 1/3 field ($B=1.8$~T) and current (5MA).  This is scenario \#1000003 in the IMAS database. A poloidal cross-section together with flux profiles is shown in Fig. \ref{fig:ITER_scen1}. We have applied steps 1 and 2 to reproduce this configuration in HELENA+ATF.

\begin{figure}[h]
\centering
\includegraphics[width=80mm]{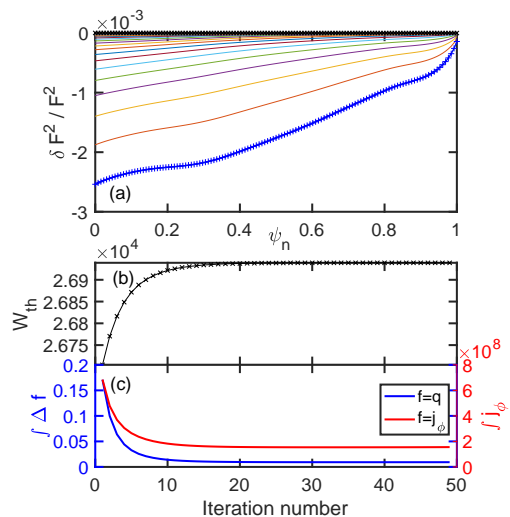}
\caption{\label{fig:iteration} 
Iteration dependancy of (a) $\delta F^2/F^2$ with $\psi_n$, showing the first iteration (blue) and last (black), 
(b) variation of $W_{th}$ and (c) $\Delta q$ and $\Delta J_\phi$ with iteration loop number}
\end{figure}

As illustration, we have next computed the equilibrium mapping for $T_\para/T_\perp = 0.8 (\code{THTOF}=0.2)$. Figure
\ref{fig:iteration} shows the variation in parameters with iteration. Panel (a) shows the change in $\delta F^2(\psi)$ profiles for different iteration numbers.  
The analysis shows $\delta F^2<0$ everywhere, and peak on-axis.  This corresponds to shifting current inward, lowering the $q$ profile. 
This is consistent with earlier work in which the unconstrained addition of anisotropy produced an increase in on-axis safety factor \cite{Hole_anisotropy_11}. 
Figure \ref{fig:iteration}(b) and \ref{fig:iteration}(c) shows the variation of stored energy $W_{th}$ and metrics $\Delta q$ and $\Delta J_\phi$ with iteration loop over Eqs. (\ref{eq:J_update}) - (\ref{eq:F_update}) for $\alpha=1, \lambda = 0.4$.  Convergence in the stored
energy is clear, as is convergence in the two metrics $\delta q$ and $\delta J_\phi$. 

\begin{figure}[h]
\centering
\includegraphics[width=80mm]{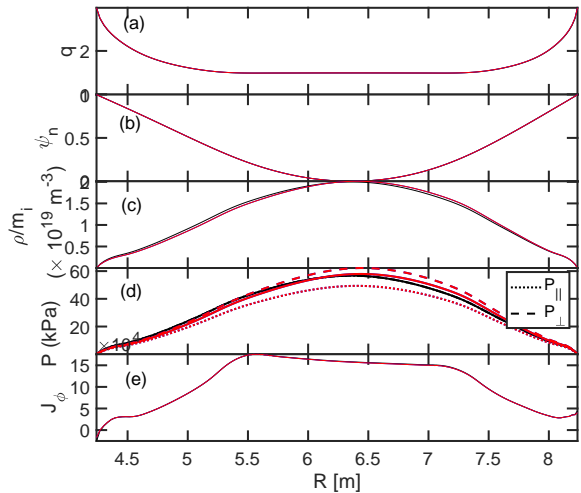}
\caption{\label{fig:ITER_anis} 
Solutions of isotropic (black) and anisotropic (red) equilibrium for $T_\para/T_\perp = 0.8$. Panels (a) through (e) show profiles of  $ q, \psi_n, \rho, P $ and $J_\phi$ respectively. The bold red line in panel (d) is the summative pressure $(2 p_\perp + p_\para)/3$.}
\end{figure}

\begin{figure}[h]
\centering

\subfigure[]{
\includegraphics[width=60mm]{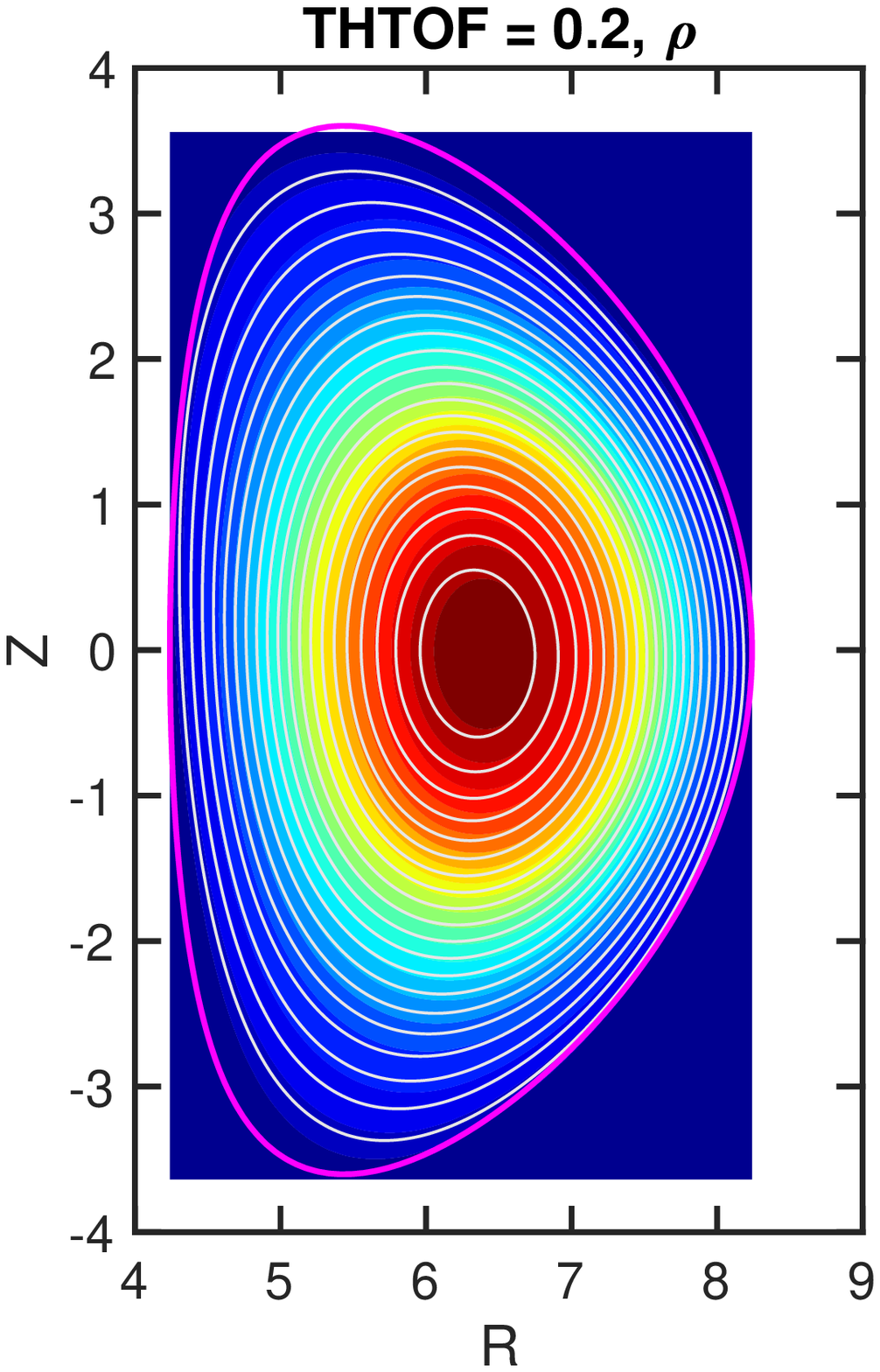}
\hspace{-2cm}}
\subfigure[]{
\includegraphics[width=60mm]{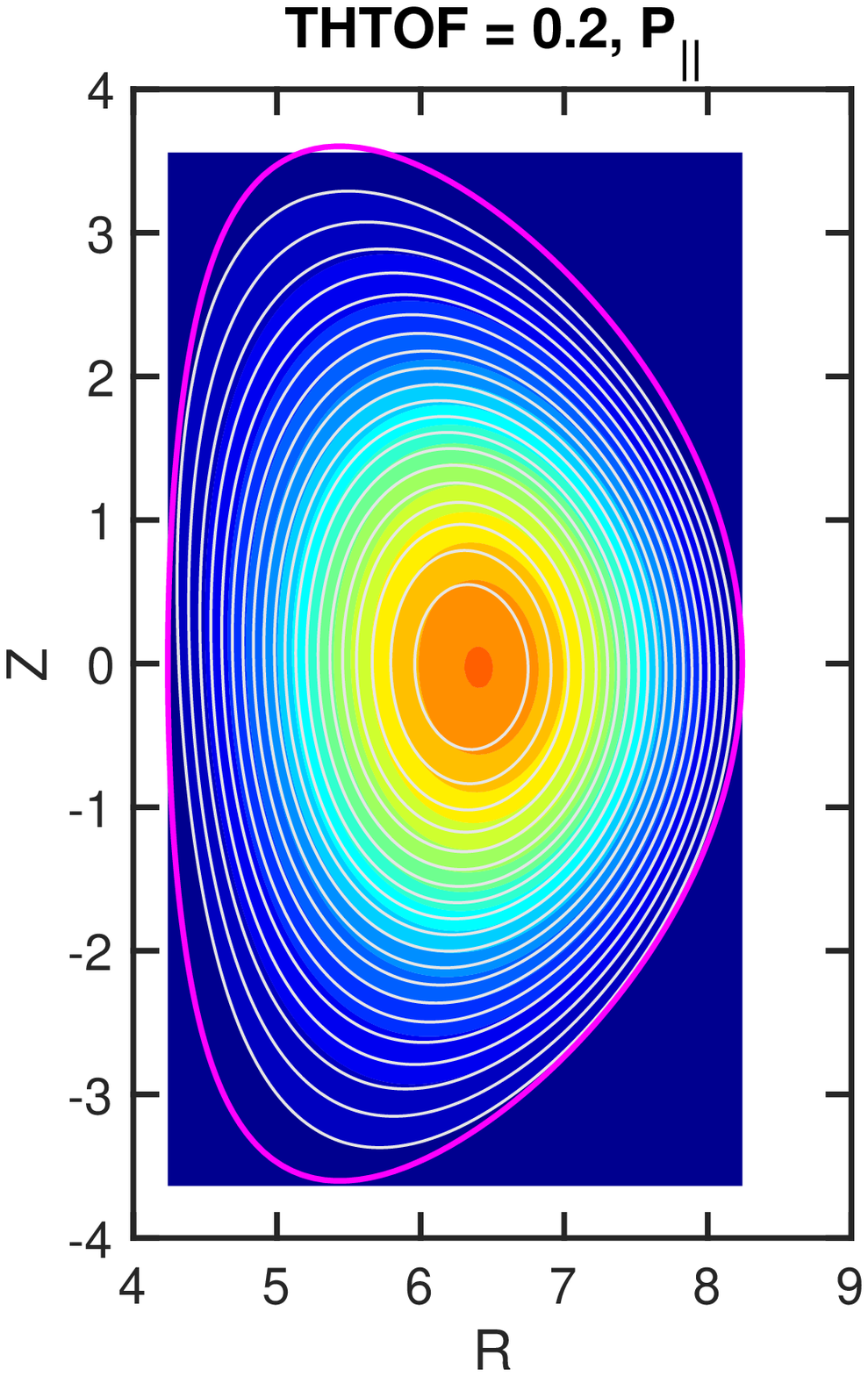}
\hspace{-2cm}}
\subfigure[]{
\includegraphics[width=60mm]{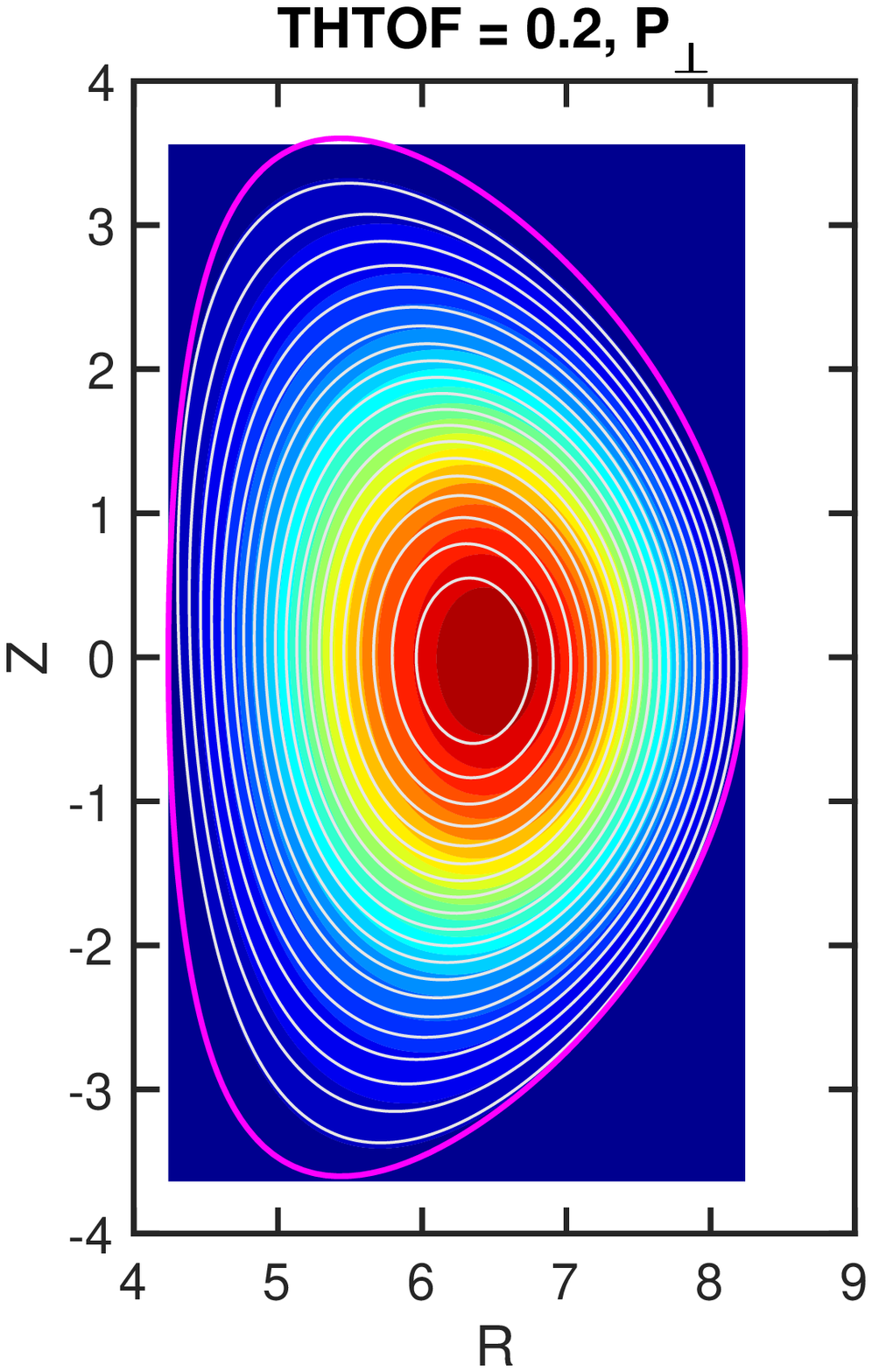}
}
\caption{\label{fig:ITER_contours} 
Plots of (a) density $\rho$, (b) parallel pressure $p_\para$ and (c) perpendicular pressure $p_\perp$ contours overlaid with 
magnetic flux surfaces (white) for the case $T_\para/T_\perp = 0.8$. }
\end{figure}

\begin{figure}[h]
\centering
\includegraphics[width=80mm]{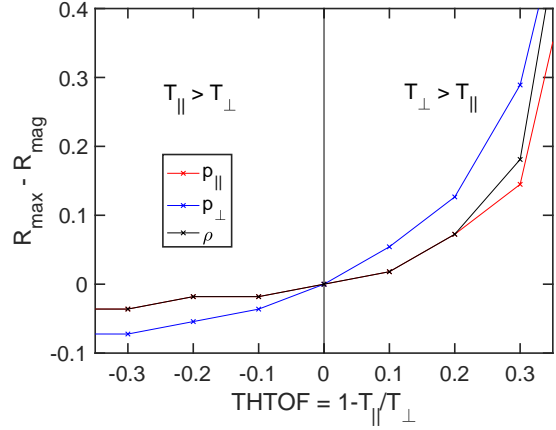}
\caption{\label{fig:anisotropy_scan} 
Calculation of density $\rho$, parallel pressure $p_\para$, and perpendicular pressure $p_\perp$ peaks  from magnetic axis as a function of $\Theta = 1 - 	T_\para/T_\perp $. }
\end{figure}

The equilibrium solution for $T_\para/T_\perp = 0.8 (\code{THTOF}=0.2)$ with mapping selection 
$\alpha =0$ is shown in Fig. \ref{fig:ITER_anis}. 
With the inclusion 
of anisotropy and rotation the physical profiles are no longer functions of flux surface.  As such, we have plotted 
quantities across a mid-plane chord.  Figures \ref{fig:ITER_anis}(a) and (b) show the $q$ profile and $\psi$ surfaces are  identical for the two cases, demonstrating that the remapping techniques have worked.  
Figure \ref{fig:ITER_anis}(c) shows the density profile is shifted outboard for the anisotropic case. For constant $T_\para$ profile the widened density profile has led to an increase in parallel and perpendicular pressure, shown in Fig.   \ref{fig:ITER_anis}(d).
Figure \ref{fig:ITER_contours} shows contours of constant density, parallel and pressure overlaid on constant flux surfaces.
The peak density, parallel and perpendicular pressure have all shifted outboard. 
This departure of pressure from flux surfaces was also observed in the analytic treatment of Cooper \etal \cite{Cooper_80}.
In that work tensor pressure equilibria were solved analytically for D-shaped cross-sections. 
Density, parallel and perpendicular pressure surfaces shifted inboard for the cases studied, for which $T_\para > T_\perp$. 
As the beam became more perpendicular the magnitude of shift decreased. 
We have performed a systematic scan of density shift as a function of $\Theta_0$, shown in Fig \ref{fig:anisotropy_scan}. 
Our results confirm the inboard shift seen in Cooper, but show significantly larger outboard shift for $T_\perp > T_\para$.

\begin{figure}[h]
\centering
\subfigure[]{
\includegraphics[width=60mm]{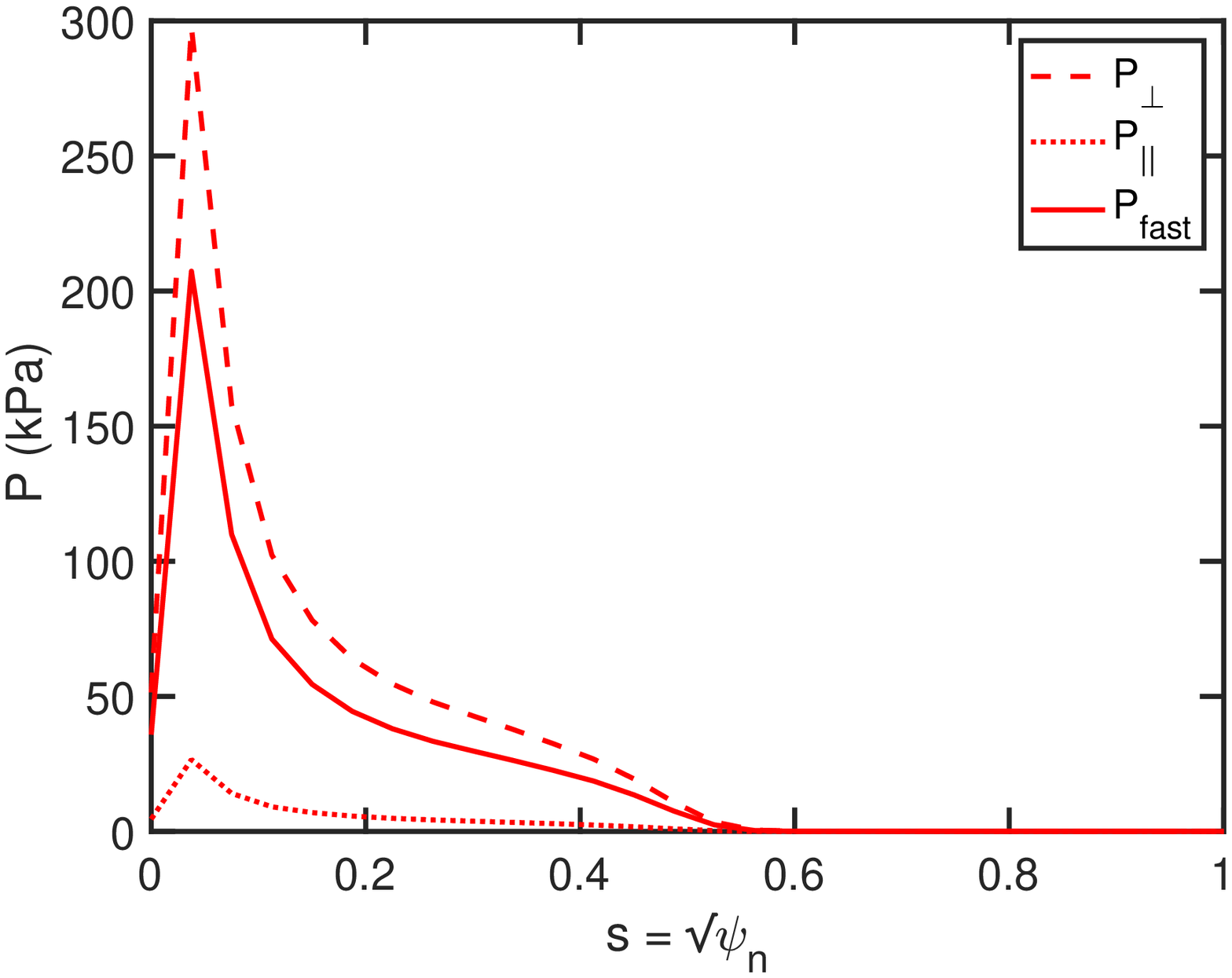}}
\subfigure[]{
\includegraphics[width=60mm]{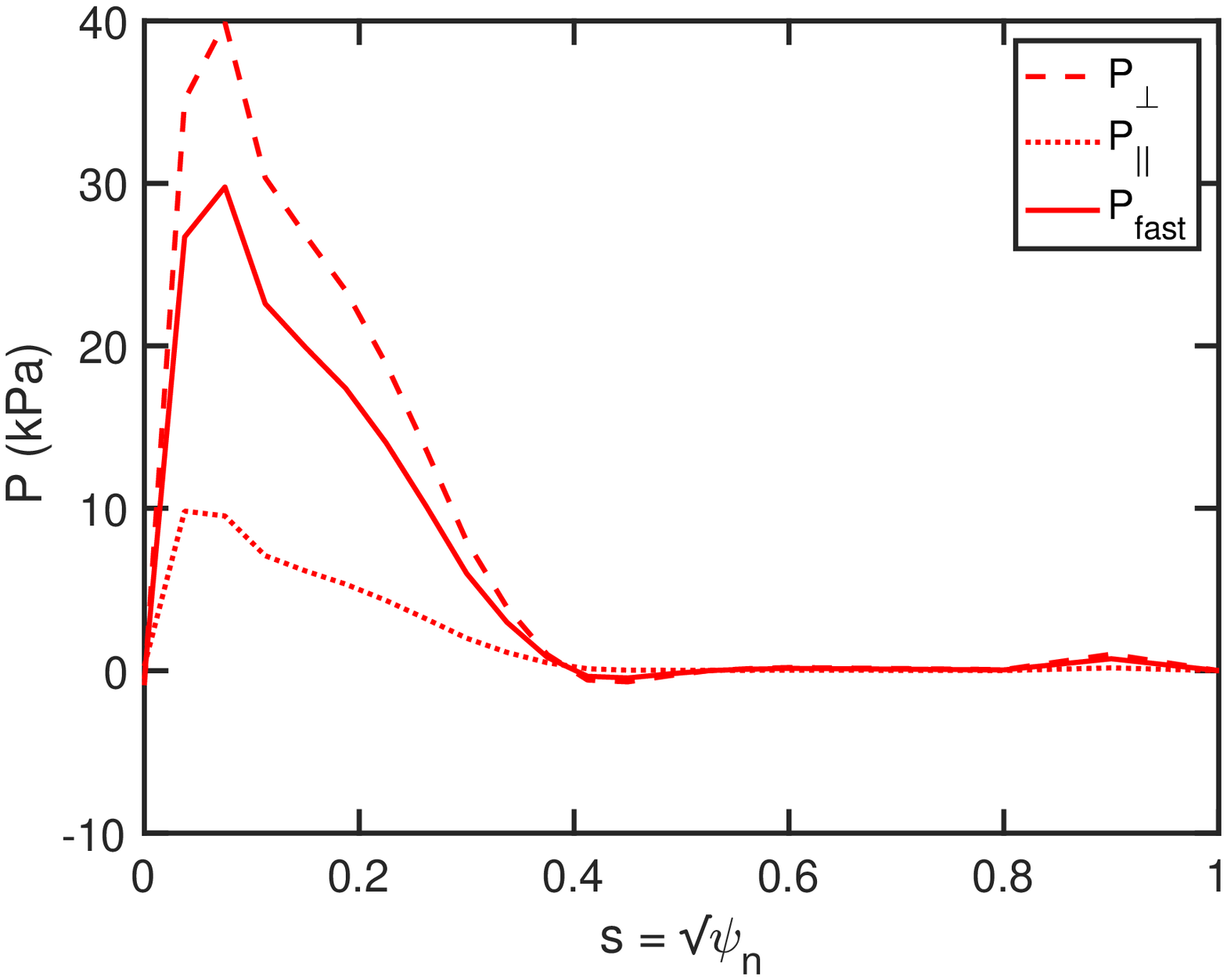}}
\caption{\label{fig:ICRH_sim} 
ICRH pressure PION simulation for (a) Case T.a of \cite{Arbina_2019} with $n_e = 0.5 n_G$ and (b) case T.b of \cite{Arbina_2019} with $n_e = 0.9 n_G$. 
In both cases B=1.74 T. These are IMAS scenarios \#100014-1 and  \#100015-1, respectively}
\end{figure}

So far, we have examined the impact of changing the central anisotropy.    The anisotropy of the distribution function can be computed using a combination of ray-tracing or global wave codes and  Fokker Planck or Monte-Carlo code. The ICRF modelling code PION \cite{Eriksson_93} uses simplified models of the power deposition and velocity distributions to provide a time resolved distribution function. It has been extensively compared against experimental results for a large variety of ICRF schemes on JET, AUG, DIII-D and Tore Supra. Recently \cite{Arbina_2019}, PION was incorporated into the ITER Integrated Modelling and Analysis Suite (IMAS) \cite{Imbeaux_2015}, and used to compute velocity distribution functions in the ITER non-activated phase. This affords computation of the moments of the fast ion distribution function give the density $n$, parallel speed $u_\para$, and parallel and perpendicular pressures:
\begin{eqnarray}
n & = & \int_0^\infty \int_{-1}^{1}  f(E, \lambda) d \lambda dE,  \label{eq:n} \\
n u_\para & = & \int_0^\infty \int_{-1}^{1} v_\para f(E, \lambda) d \lambda dE,  \label{eq:u_para} \\
p_\para & = & m \int_0^\infty \int_{-1}^{1} (v_\para - u_\para)^2 f(E, \lambda) d \lambda dE,  \label{eq:p_para} \\
p_\perp & = & \frac{m}{2} \int_0^\infty \int_{-1}^{1} v_\perp^2 f(E, \lambda) d \lambda dE, \label{eq:p_perp} \\
\end{eqnarray}  
where $E$ is the energy and $\lambda = v_\para/v_\perp$ the pitch angle. 
We have used the velocity distribution functions from PION, computed moments to obtain fast ion $P_\para$ and $P_\perp$ as a function of flux surface, summed the thermal and ion resonant populations to obtain a measure of the total anisotropic pressure. Our focus case is T.a and T.b in Arbina \etal \cite{Arbina_2019}, which are IMAS scenarios \#100014-1 and \#100015-1, respectively.  These both correspond to 1/3 field strength $B=1.74$~T, with $n_e=0.5 n_G$ and $n_e = 0.9 n_G$, with $n_G$ the Greenwald density.    Figure \ref{fig:ICRH_sim} shows the computed pressure components as a function of poloidal flux.  Here we have used the kinetic energy closure 
$p_{fast} = (2 p_\perp + p_\para)/3$. 

\begin{figure}[h]
\centering
\includegraphics[width=60mm]{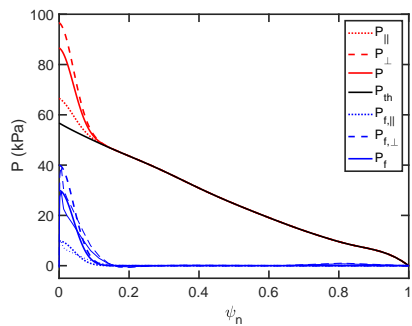}
\caption{\label{fig:ICRH_tot} 
Fitted and total pressure for case T.b of  \cite{Arbina_2019} with $n_e = 0.9 n_G$. The ICRH simulated data from PION is light, and the fitted 
profiles heavy.}
\end{figure}

\begin{figure}[h]
\centering
\includegraphics[width=80mm]{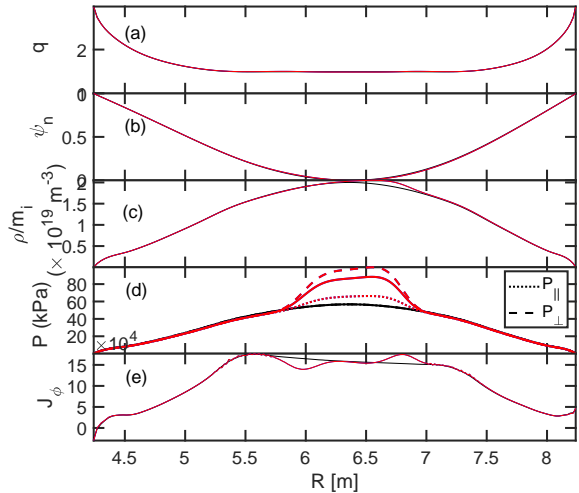}
\caption{\label{fig:ITER_ICRH} 
Solutions of isotropic (black) and anisotropic (red) equilibrium for the ICRH pressure derived from case T.b of \cite{Arbina_2019}. Panels (a) through (e) show profiles of  $ q, \psi_n, \rho, P $ and $J_\phi$ respectively.}
\end{figure}

To enable a study of the impact of anisotropy for case T.b we have fitted Gaussian profiles to the ICRH pressure components, with mean selected on-axis at $\psi_n=0$. This is shown in Fig. \ref{fig:ICRH_tot}, in which the fast ion components are shown.  Also shown is the thermal pressure profile for the simulation, and finally the summative anisotropic pressure.  The summative anisotropic pressure has been used in the remapping steps in Sec. 2. to construct the ITER scenario with anisotropic pressure, as shown in  Fig. \ref{fig:ITER_ICRH}.

Finally, we have implemented the remapping steps in Sec. 2 to case T.b giving the equilibrium solution shown in Fig. \ref{fig:ITER_ICRH}. 
An immediate feature of the solution is that the fast ion pressure in the plasma core has displaced the pressure and density off-axis. Indeed, 
the density peak has shifted to R=6.56m, an outboard shift of 0.16m from the magnetic axis.  A change in the toroidal current profile adjacent to the magnetic axis is also visible: this arises from step 4 of the equilibrium remapping to compensate the addition of the ICRH pressure gradient term.

\section{MHD Continuum modes}

\begin{figure}[h]
\centering
\includegraphics[width=80mm]{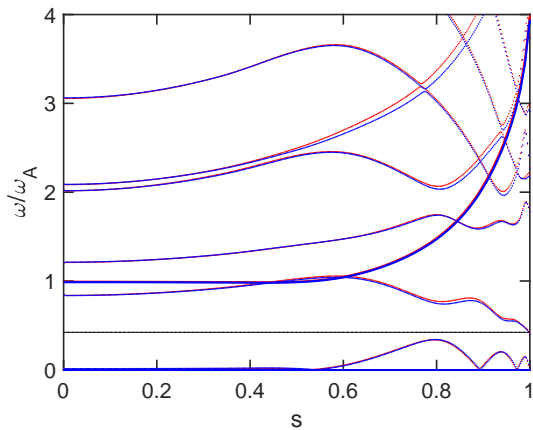}
\caption{\label{fig:ITER_shear_n1} 
Shear Alfv\'{e}n continuum for $n=1$ with $T_\para/T_\perp = 0.8$ (blue) and $T_\para/T_\perp = 1.2$ (red). Also shown is the TAE frequency (black).
The case for which  $T_\para = T_\perp$ is IMAS scenario \#1000003.}
\end{figure}

\begin{figure}[h]
\centering
\includegraphics[width=80mm]{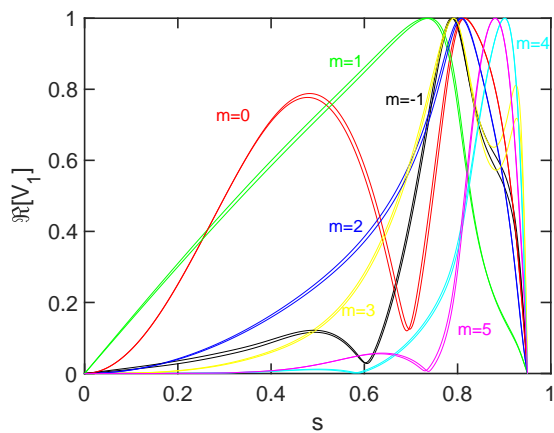}
\caption{\label{fig:ITER_shear_n1_modes} 
Overplotted TAE modes for $\omega/\omega_A = 0.4226$ for $T_\para / T_\perp = 0.8$, and $\omega/\omega_A = 0.4234$ for $T_\para / T_\perp = 1.2$.  }
\end{figure}

As a first step to exploring the mode spectrum we have computed the shear Alfven (incompressional) and compressional continuum for isotropic and anisotropic plasmas using \code{CSMIS\_AD}, which is the continuum solver in the anisotropic MHD stability code \code{MISHKA-AD} \cite{Qu_2015}. Figure \ref{fig:ITER_shear_n1} overplots the $n=1$ shear Alfv\'{e}n continuum for $T_\para/T_\perp = 0.8$ and $T_\para/T_\perp = 1.2$.  The shear Alfven continuum are virtually identical for the two cases. We have examined each solution for a candidate global mode,  and found an $n=1$ TAE global mode at $\omega/\omega_A = 0.4226$ for $T_\para / T_\perp = 0.8$, and $\omega/\omega_A = 0.4234$ for $T_\para / T_\perp = 1.2$.
Figure \ref{fig:ITER_shear_n1_modes} overplots the eigenfunction of the two $n=1$ TAE modes. The mode structure is identical. Our conclusion is that the shear Alfv\'{e}n branch, mode frequencies, mode structure is unaffected by anisotropy.  As the magnetic geometry for the two cases is identical and the frequency the same the resonance condition for particles will also be identical, and hence the linear mode drive and ensuing nonlinear dynamics will be the same. Our analysis complements recent work by Gorelenkov and Zakharov, who added the impact of fast ion orbit width to a pressure anisotropy and toroidal flow formulation. They showed that the inclusion of finite orbit width effects reduces the Shafranov shift with increasing beam injection energy. Consequently, the upshift in TAE frequency due to the Shafranov shift decreases and TAEs experience reduced continuum damping.  

\begin{figure}[h]
\centering
\includegraphics[width=80mm]{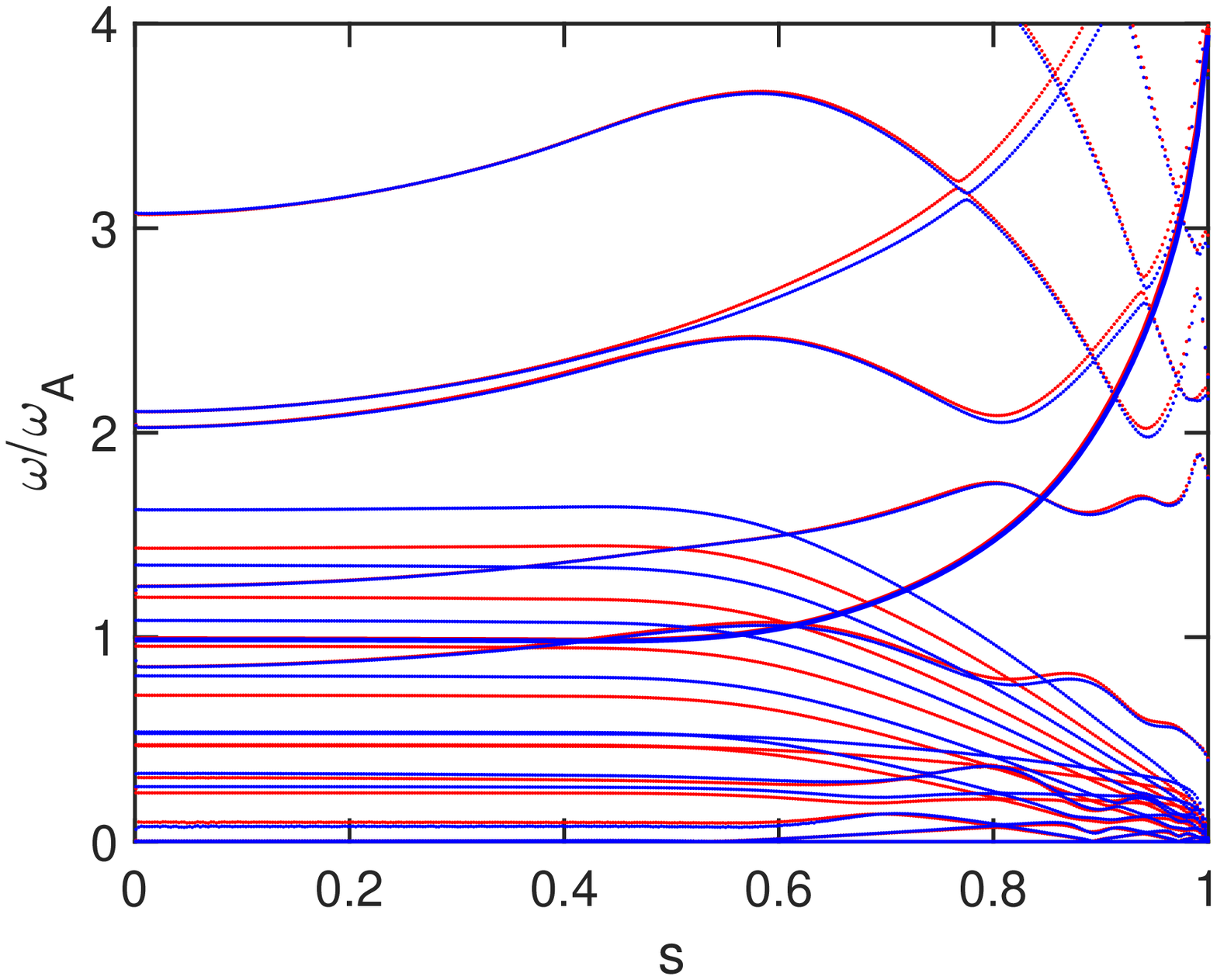}
\caption{\label{fig:ITER_comp_n1} 
Compressional Alfv\'{e}n continuum for $n=1$ with $T_\para/T_\perp = 0.8$ (blue) and $T_\para/T_\perp = 1.2$ (red).  }
\end{figure}

We have also computed the compressional branch, with the adiabatic index $\gamma = 5/3$. Figure \ref{fig:ITER_comp_n1} shows the $n=1$ continuum.  In contrast to the shear continuum, significant shift is evident in the Alfven-acoustic continua.  This suggests modification to the compressional beta-induced Alfven eigenmodes  (BAEs) and beta-induced Acoustic Alfven eigenmodes (BAAEs). 

\section{Conclusions}
In this work we have computed the impact of anisotropy to ITER pre-fusion power operation scenarios. To undertake this calculation a series of remapping techniques were developed to resolve the impact of pressure anisotropy as compared to the change in magnetic field configuration. Scans were conducted as a function of varying central anisotropy. The impact of anisotropy was to shift the density and pressure contours from flux surfaces: we found an outboard (inboard) shift of peak density for $T_\para > T_\perp ( T_\para < T_\perp)$. Computation for a realistic ITER scenario with ICRH computed fast ions has been undertaken. This showed the impact of dominantly core anisotropy with $P_\perp > P_\para$ produces an outboard shift of the density 0.16~m off-axis.
In separate working we have found demonstrated that the impact of anisotropy on the shear (compressional) continuum is small (large).  For the shear branch, we demonstrated that Alfven gap modes do not change in frequency or mode structure. In contrast, the frequencies of the compressional branch of the continuum are substantively changed.  This suggests modification to BAE and BAAE modes.  

Our results suggest a number of directions for further research.  The shift in density is purely due to pressure anisotropy, not flow.  It would be useful to explore experimentally whether shift from the magnetic axis can be resolved between toroidal flow and anisotropy. We have undertaken a study of only one ITER scenario with ICRH.  A more complete study over the full range of proposed ITER scenarios, with both ICRH and NBI would be informative. We have found demonstrated that the impact of anisotropy on the shear (compressional) continuum is small (large). The former suggests that the stability of Alfv\'{e}n gap modes is unaffected by anisotropy, while the frequency and stability of compressional modes such as the BAE and BAAE could be affected. Experimentally, the latter could be investigated by examining the impact of varying external heating on compressional modes. Confirmation that the shear branch is unaltered
could be performed by undertaking several experiments with similar $q$ profiles but different heating.

\section*{Acknowledgments}
This work was partly funded by the Australian Government through Australian Research Council grants DP140100790 and DP170102606.
The authors gratefully acknowledge the support of G. A. Huysmans from CEA Cadarache in the provision of CSCAS, HELENA and MISHKA codes.
The views and opinions expressed herein do not necessarily reflect those of the ITER Organization.
This work has been partly carried out within the framework of the EUROfusion Consortium and has received funding from the Euratom research and training programme 2014-2018 and 2019-2020 under grant agreement No 633053. The views and opinions expressed herein do not necessarily reflect those of the European Commission.

\section*{References}
 
\bibliographystyle{unsrt}
\bibliography{MHD_energ_particles}

\begin{thebibliography}{10}

\bibitem{Hole_anisotropy_11}
M.~J. Hole, G.~von Nessi, M.~Fitzgerald, K.~G. McClements, J.~Svensson, and the
  MAST~team.
\newblock Identifying the impact of rotation, anisotropy, and energetic
  particle physics in tokamaks.
\newblock {\em Plasma Phys. Control. Fusion}, 53:074021, 2011.

\bibitem{Zwingmann_01}
W.~Zwingmann, L.~G. Eriksson, and P.~Stubber.
\newblock Equilibrium analysis of tokamak discharges with anisotropic pressure.
\newblock {\em Plasma Phys. Control. Fusion}, 2001.

\bibitem{Cooper_80}
W.~A. Cooper, G.~Bateman, D.~B. Nelson, and T.~Kammash.
\newblock {Beam-induced tensor pressure tokamak equilibria}.
\newblock {\em Nucl. Fus.}, 20(8):985--992, 1980.

\bibitem{Salberta_1987}
E.~R. Salberta, R.~C. Grimm, J.~L. Johnson, J.~Manickam, and W.~N. Tang.
\newblock { Anisotropic pressure tokamak equilibrium and stability
  considerations}.
\newblock {\em Phys. Fluids}, 30:2796--2805, 1987.

\bibitem{Iacono_90}
R.~Iacono, A.~Bonderson, F.~Troyon, and R.~Gruber.
\newblock {Axisymmetric toroidal equilibrium with flow and anisotropic
  pressure}.
\newblock {\em Phys. Fluids B}, 2(8):1794--1803, 1990.

\bibitem{Ilgisonis_1996}
V.~I. Ilgisonis.
\newblock {Anisotropic plasma with flows in tokamak: Steady state and
  stability}.
\newblock {\em Phys. Plas.}, 3:4577, 1996.

\bibitem{Guazzotto_04}
L.~Guazzotto, R.~Betti, J.~Manickam, and S.~Kaye.
\newblock {Numerical study of tokamak equilibria with arbitrary flow}.
\newblock {\em Phys. Plasmas}, 11(2):604--614, February 2004.

\bibitem{Hole_Dennis_09}
M.~J. Hole and G.~Dennis.
\newblock {Energetically resolved multiple-fluid equilibrium of tokamak
  plasmas}.
\newblock {\em Plas. Phys. Con. Fus.}, 51:035014, 2009.

\bibitem{Fitzgerald_2013}
M.~Fitzgerald, L.~C. Appel, and M.~J. Hole.
\newblock {EFIT tokamak equilibria with toroidal flow and anisotropic pressure
  using the two-temperature guiding-centre plasma}.
\newblock {\em Nuc. Fus.}, page accepted 01/10/2013, 2013.

\bibitem{Qu_2014}
Z~S Qu, M~Fitzgerald, and M~J Hole.
\newblock {Analysing the impact of anisotropy pressure on tokamak equilibria}.
\newblock {\em Plasma Phys. Control. Fusion}, 56:075007, 2014.

\bibitem{Gorelenkov_2018}
N.~N. Gorelenkov and L.~E. Zakharov.
\newblock {Plasma equilibrium with fast ion orbit width, pressure anisotropy,
  and toroidal flow effects}.
\newblock {\em Nuc. Fus.s}, 58:082031, 2018.

\bibitem{Hole_2013}
M~J Hole, G~von Nessi, M~Fitzgerald, and the MAST~team.
\newblock {Fast particle modifications to equilibria and resulting changes to
  Alfv\'{e}n wave modes in tokamaks}.
\newblock {\em Plasma Phys. Control. Fusion}, 55:014007, 2013.

\bibitem{CSCAS_Poedts_93}
Poedts and Schwartz.
\newblock {Computation of the ideal-MHD Continuous Spectrum in Axisymmetric
  Plasmas}.
\newblock {\em Jl. Computational Physics}, 105:165--168, 1993.

\bibitem{MISHKA1_97}
A.~B. Mikhailovskii, G.~T.~A. Huysmans, W.~O.~K. Kerner, and S.~E. Sharapov.
\newblock {Optimization of Computational MHD Normal-Mode Analysis for
  Tokamaks}.
\newblock {\em Plas. Phys. Rep.}, 21(10):844--857, 1997.

\bibitem{Layden_16}
B.~Layden, Z.S. Qu, M.~Fitzgerald, and M.J. Hole.
\newblock {Impact of pressure anisotropy on magnetic configuration and
  stability}.
\newblock {\em Nucl. Fusion}, 56, 2016.

\bibitem{Eriksson_93}
L-G. Eriksson, T.~Hellsten, and U.~Will\'{e}n.
\newblock {Comparison of Time Dependent Simulations with Experiments in Ion
  Cyclotron Heated Plasmas }.
\newblock {\em Nuc. Fus.}, 33(7):1037--1048, 1993.

\bibitem{Arbina_2019}
I.L. Arbina, M.J. Mantsinen, X.~S\'{a}ez, D.~Gallart, A.~Guti\'{e}rrez,
  D.~Taylor, T.~Johnson, S.D. Pinches, M.~Schneider, and the
  EUROfusion-IM~Team.
\newblock {First applications of the ICRF modelling code PION in the ITER
  Integrated Modelling and Analysis Suite}.
\newblock In {\em Proc. 36th EPS Conf. Controlled Fusion Plas. Physics}, page
  P4.1079, Milan, Itlay, July 2019.

\bibitem{Imbeaux_2015}
F.~Imbeaux, S.D. Pinches, J.B. Lister, Y.~Buravand, T.~Casper, B.~Duval,
  B.Guillerminet, M.~Hosokawa, W.~Houlberg, P.~Huynh, S.H. Kim, G.~Manduchi,
  M.Owsiak, B.~Palak, M.~Plociennik, G.~Rouault, O.~Sauter, and P.~Strand.
\newblock {Design and first applications of the ITER integrated modelling and
  analysis suite}.
\newblock {\em Nuc. Fus.}, 55:123006, 2015.

\bibitem{Qu_2015}
Z.~S. Qu, M.~Fitzgerald, and M.~J. Hole.
\newblock {Modeling the effect of anisotropic pressure on tokamak plasmas
  normal modes and continuum using fluid approaches}.
\newblock {\em Plasma Phys. Control. Fusion}, 57:095005, 2015.

\end{thebibliography}

\end{document}